# Introduction to the Special Issue on Quantum Cognition

Peter Bruza, Jerome Busemeyer, and Liane Gabora

The subject of this special issue is quantum models of cognition. At first sight it may seem bizarre, even ridiculous, to draw a connection between quantum mechanics, a highly successful theory usually understood as modeling sub-atomic phenomena, and cognitive science. However, a growing number of researchers are looking to quantum theory to circumvent stubborn problems within their own fields. This is also true within cognitive science and related areas, hence this special issue.

## Justification

Given the nascent state of this field, some words of justification are warranted. The researchers just mentioned are not concerned with modeling physical phenomena, but instead turn to quantum theory as a fresh conceptual framework in which to consider problems, as well as a source of alternative formal tools. With less than perfect accuracy, there are two aspects of quantum theory which open the door to addressing problems in a totally new light (see Barros & Suppes on this issue). The first is known as "contextuality". Even within quantum theory, contextuality is a subtle notion to grasp. However, one way to understand it is in terms of interference. When quantum systems are in superposed states, they can interfere. Interference has an expression within the underlying probabilistic apparatus, which is not to be found within the "classical" probabilistic framework.

The second aspect is "quantum entanglement". Entanglement is a bizarre phenomenon in which seemingly separated quantum systems behave as one. Entanglement has led to ongoing philosophical debate about the nature of reality, and was at the heart of a famous debate between Neils Bohr and Albert Einstein. Erwin Schrödinger, one of the founding fathers of quantum theory wrote in 1935, "I would not call [entanglement] *one* but rather *the* characteristic trait of quantum mechanics, the one that enforces its entire departure from classical lines of thought." In quantum physics, the term "non-locality" is often mentioned in conjunction with entanglement, meaning that performing a measurement on one system directly and instantaneously affects the state another system, even when such systems are separated. The key move made by researchers exploiting notions of entanglement outside of physics is to view it as a promising means to model *non-separability* of cognitive states.

Several phenomena in psychology that have stubbornly resisted traditional modeling techniques are showing promise with a new modeling approach inspired by the formalisms of quantum mechanics. Before going into any details about this new approach, let us look at a couple of these phenomena:

*Decision making* – When people are given a chance to play a particular gamble twice, if they think they won the first play, or alternatively if they think they lost the first play, then the majority chooses to play again on the second round. Given these preferences, they should also play the second round even if they don't think about the outcome of the first round. Yet people do just the opposite in the latter case (Tversky & Shafir, 1992). This finding violates the law of total probability, yet it can be explained as a quantum interference effect in a manner similar to the explanation for the results from two-hole experiments in physics (Pothos & Busemeyer, 2009).

*The Contextual Nature* of *Concept*s and their *Combinations* –When quantum entities become *entangled*, they form a new entity with properties different from either constituent, and one cannot manipulate one constituent without simultaneously affecting the other. The mathematics of entanglement has been used to model the nonmonotonic relations observed among concepts when they are combined to form a new concepts such as STONE LION (Gabora & Aerts, 2005).

**The Legacy and Contents of this Special Issue**

Quantum theory was originally invented by physicists to explain findings that seemed paradoxical from a standard physical view point. Later Von Neumann (1932) provided an axiomatic foundation for quantum theory, and by doing so, he discovered that it implied a new type of logic and probability theory. Consequently, there are now two general theories for assigning probabilities to events: classical (Kolmogorov, 1933) and quantum (von Neumann, 1933). Classic probability theory defines events as subsets of a universal set, which obey all the laws of Boolean algebra. Quantum theory defines events as subspaces of a Hilbert space, which obey all the laws of Boolean algebra *except* the distributive axiom. Following from the distributive axiom, classic probability theory adheres to one of its most important theorems, the law of total probability. In contrast, because quantum logic does not have to obey the distributive law, quantum probabilities do not have to obey the law of total probability. Based on a Hilbert space representation, quantum probabilities are required to obey two other laws: reciprocity and double stochasticity. Classic probability theory is not based on a Hilbert space representation, and so it does not have to obey the latter two laws. Thus there is a substantial disagreement between the two probability theories, and which theory is best is an empirical question.

There is a short, but significant history of applying the formalisms of quantum theory to topics in psychology. Initially, actual quantum physics was used to model the brain and explain consciousness (Hameroff, 1994, 1998; Hameroff & Penrose, 1996; Jibu et al., 1994; Pribram, 1991; Penrose, 1993). An influential web-based course on quantum consciousness was hosted by the University of Arizona in Tucson, as well as series of international conferences on the topic. At about the same time, ideas for applying quantum formalisms to cognition appeared (Aerts & Aerts, 1994; Atmanspacher, 1992; Bordley, 1998; Khrennikov, 1999; Turvey & Shaw, 1995). Shortly afterwards, there appeared some more detailed efforts (Atmanspacher, Spilk, & Romer, 2004; Bruza & Cole, 2005; Busemeyer, Wang, & Townsend, 2006; Gabora & Aerts, 2002; Ivancevic, & Aidman, 2007; van Rijsbergen, 2004; Widdows, 2003). However, it wasn't until the first Quantum Interaction workshop at Stanford in 2007 organized by Peter Bruza, William Lawless, C. J. van Rijsbergen, and Don Sofge as part of the AAAI Spring Symposium that a community began to emerge. This first workshop was followed by workshops at Oxford (England) in 2008, Vaexjo (Sweeden) in 2008, and Saarbruken (Germany) in 2009. Tutorials also were presented annually beginning in 2008 until present at the meeting of the Cognitive Science Society.

The papers in this issue are highly interdisciplinary; their authors are based in psychology, mathematics, physics, and computer science. Several of them are speculative, as perhaps should be the case when a field is very new. Within this special issue, interference is being exploited by a number of researchers, for example, in relation to new models of human judgment and decision making (Busemeyer, Wang, & Lampert-Mogiliansky, Franco, LaMura, Narens, Khrennikov & Haven, Lampert-Mogiliansky, Zamir, & Zwirn). Entanglement is being explored for word associates in human memory (Bruza, Kitto, Nelson and McEvoy), modeling words as non-separable in concept combinations, and in relation to emergent concepts (Aerts, Czachor, de Moor & Aerts) and emergent worldviews (Gabora and Aerts). We believe these papers pave the way for a very promising new direction of psychological investigation.